\documentclass[aps,twocolumn,showpacs,preprintnumbers]{revtex4}

\usepackage{graphicx,umlaut}

\begin{document}

\title{Three-dimensional MHD simulation of expanding magnetic flux ropes}

\author{ L. Arnold$^{1}$,
J. Dreher$^{1}$,
R. Grauer$^{1}$,
H. Soltwisch$^{2}$ and  
H. Stein$^{2}$\\
$^{1}$ Theoretische Physik I, 
Ruhr-Universit\"at, 44780 Bochum, Germany \\
$^{2}$ Experimentalphysik V, 
Ruhr-Universit\"at, 44780 Bochum, Germany \\
}

\newcommand{\vv}
{\vec{v}}
\newcommand{\ve}
{\vec{e}}
\newcommand{\vx}
{\vec{x}}
\newcommand{\vB}
{\vec{B}}
\newcommand{\vA}
{\vec{A}}
\newcommand{\vj}
{\vec{j}}
\newcommand{\vm}
{\vec{m}}

\newcommand{\dBdt}
{\frac{\partial\vec{B}}{\partial t}}
\newcommand{\drhodt}
{\frac{\partial\rho}{\partial t}}
\newcommand{\dvdt}
{\frac{\partial\vec{v}}{\partial t}}
\newcommand{\dpsidt}
{\frac{\partial\psi}{\partial t}}

\begin{abstract}

Three-dimensional, time-dependent numerical simulations of the dynamics
of magnetic flux ropes are presented.  The simulations are targeted
towards an experiment previously conducted at CalTech (Bellan, P. M.
and J. F. Hansen, Phys. Plasmas, {\bf 5}, 1991 (1998)) which aimed at
simulating Solar prominence eruptions in the laboratory.  
The plasma dynamics is described by ideal MHD using different models for
the evolution of the mass density.
Key features of the reported experimental observations like pinching
of the current loop, its expansion and distortion into helical shape are
reproduced in the numerical simulations.
Details of the final structure depend on the choice of a
specific model for the mass density.
\end{abstract}


\maketitle

\section{Introduction}

Magnetic fields play an important role for the structure and the
dynamical behavior of the solar atmosphere. 
Well-known examples for structural features found on a variety of
spatial scales are helmet streamers \citep[][]{bib:wiegelmann},
sunspots, coronal loops and filaments, while flares, loop eruptions
and coronal mass ejections
are dynamical phenomena related to magnetic energy release  \citep[][]{bib:aschwanden}.
The importance of the magnetic field structure and its evolution has
prompted a lot of theoretical investigation in the past. For example,
T\"or\"ok et al. \citep[][]{bib:kliem} use numerical simulations to study
a scenario for loop eruptions due to kink mode instabilities, using
the loop model by Titov and Démoulin \citep[][]{bib:titov-demoulin}.

An entirely different approach to study the evolution and creation of
magnetic signatures found in the solar corona is the use of
laboratory experiments.
Bellan and Hansen report from an experiment
which is considered a laboratory model for Solar prominence eruptions
\cite[][]{bib:bellan2000,bib:bellan2001},
and a similar experiment (``FlareLab'') has recently been set up at
Ruhr-Universit\"at Bochum.
In the latter, the long term goal is to study a variety of magnetic field
configurations and to employ extended plasma diagnostics.

\begin{figure}[t!]
  \begin{center}
    \includegraphics[width=1.0\columnwidth]{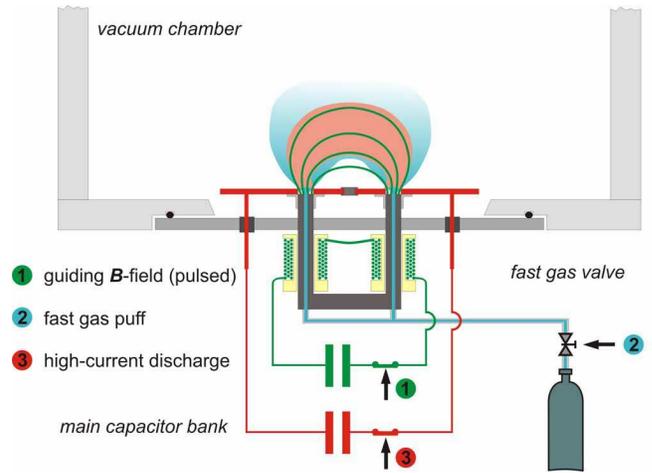}
  \end{center}
\caption{\label{fig:exp-scheme} Sketch of the initial experimental
setup used by Bellan \citep[][]{bib:bellan2001} and in ``FlareLab'' at
Bochum Universit\"at.
}
\end{figure}

\begin{figure}[t!]
  \begin{center}
    \includegraphics[width=0.9\columnwidth]{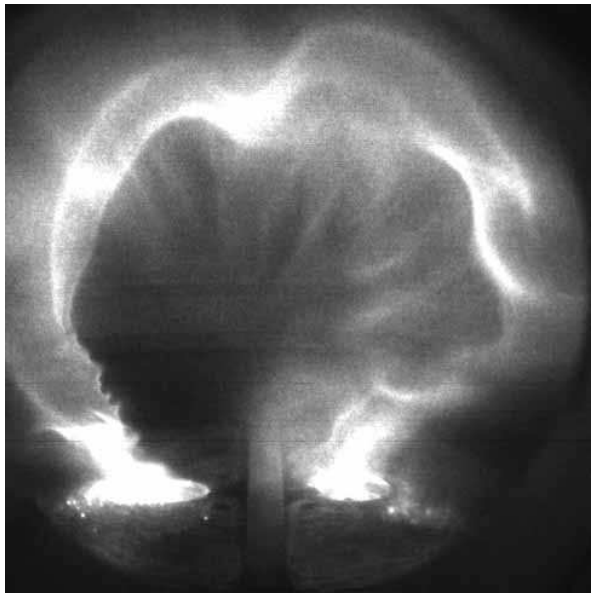}
  \end{center}
\caption{\label{fig:flarelab} Image taken from the ``FlareLab'' experiment
at $7\mu s$ after ignition.  The plasma arc has already expanded to
approximately three times its initial radius and helical distortions are
clearly visible.}
\end{figure}

Fig. \ref{fig:exp-scheme} shows a crude sketch of these two experiments:
A horse shoe magnet is mounted below the  bottom plate of a vacuum chamber,
producing an arc-shaped field inside the vessel.
Two electrodes in the vicinity of the magnet poles are connected to a
capacitor bank.
In the experiment, hydrogen gas is puffed into the chamber shortly
before the capacitor voltage is connected to the electrodes.
A plasma arc is created by ionisation of the hydrogen, causing the
voltage to break down and the capacitors to be discharged.
The evolution of this arc is then followed by means of a fast camera in order to
study the dynamics that follow from the internal magnetic forces.
Images from different stages of the discharge process as published in
\citep[][]{bib:bellan2000} indicate that the arc pinches in
cross-section and expands as a whole on the scale of a few microseconds.
At a later stage, it gets further deformed and assumes a helical shape.
A corresponding photograph recently taken from ``FlareLab'' is shown in
Fig. \ref{fig:flarelab}.

Later, this basic experimental setup has been modified by Bellan et al. to
include an additional magnetic field component with field lines that
cover the entire filament structure at a right angle. 
Using this modified setup, it was
demonstrated that the ``strapping'' field component can inhibit the
arc expansion \citep[][]{bib:bellan2001a}.
Also, the interaction between two current arcs has been studied in an
other extension of the basic configuration \citep[][]{bib:bellan2arcs}.

This experimental approach offers an attractive way to gain insight
into the dynamical behavior of  magnetic structures, because it
opens a way to analyze details of the dynamics through the measurement
of physical quantities like the magnetic field, which are not directly
accessible in the solar context.
Apart from this, the possibility to control and selectively modify
parameters like boundary- and initial conditions allows systematic
investigations and the test of theoretical models or comparison with computer
simulations.
However, a number of questions arise concerning the interpretation of
the experimental observations and its relevance for solar physics:
First of all, it is not clear if the model of magneto-hydrodynamics
(MHD), which seems appropriate in the solar context and has been
adopted for the interpretation of experimental findings
\citep{bib:bellan2000}, can be applied to the experiment without
limitations.
While the plasma arc itself may be highly ionized, there is no
indicaction available from the experiment concerning the electron
density off the regions of highest luminosity. Therefore, it is
unclear if, for instance, the ``frozen-flux'' principle applies in
those surrounding regions and what the consequences of any
``non-ideal'' behavior would be.

In an attempt to obtain a better understanding for the dynamical
evolution of the plasma arc, we have carried out a number of computations
which aim at simulating the plasma dynamics in the experiment.
Using the model of ideal MHD, we can address the question whether
experimental results in fact are compatible with that model, and to
which extent it is capable of reproducing the observed structures.
In addition, detailed data is available about relevant quantities like
the magnetic field, electric current and plasma flow.
It should be stressed that these simulations are directed at the
laboratory experiment alone, and no attempt is made at this point to
draw conclusions for prominence dynamics in the solar atmosphere.

\section{Model and Numerical Treatment}
We employ the model of ideal MHD,
\begin{eqnarray}
\label{eqn:MHD-cont}\drhodt &=& -\nabla\cdot \left(\rho\vv \right) + S\\
\label{eqn:MHD-motion}\dvdt &=& -\left(\vv\cdot\nabla\right)\vv + \frac{\vj\times\vB}{\rho} - \frac{\nabla p}{\rho}
+ \nu\Delta\vv \\
\label{eqn:MHD-induction}\dBdt &=& \nabla\times\left(\vv\times\vB\right)  \\
\label{eqn:MHD-ampere} \nabla\times\vec{B} &=& \mu_0 \vec{j}.
\end{eqnarray}
Here, $\rho$, $\vv$, $\vB$, $p$ and $\vj$ denote the mass
density, plasma bulk velocity, magnetic field, thermal pressure and
electric current density, respectively.  A homogeneous kinematic
viscosity, $\nu$, is included for reasons of numerical stability.  In
(\ref{eqn:MHD-cont}), a yet unspecified source term $S$ allows the model
to deviate from plasma mass conservation. 
As described below in detail, this term is used in order to implement
different models for the plasma density in our simulations, namely i)
mass conservation, $S=0$, ii) a homogeneous density, $\rho=\rho_0$
iii) a fixed Alfv\'en velocity, $\rho\propto |\vB|^2$ and iv) a crude
model for ionization and recombination (case 4 below).  These models
are only partly motivated on physical grounds, but we note that
Eq. (\ref{eqn:MHD-motion}) is written in non-conservative form and
hence remains consistent with (\ref{eqn:MHD-cont}) even for $S\ne 0$
if it is assumed that plasma created according to $S$ has the same
velocity as the existing plasma population.
For an ionization term, this is the case if the neutral gas component
moves at the same velocity $\vec{v}$ as the charged species.
To close Eqs. (\ref{eqn:MHD-cont})--(\ref{eqn:MHD-induction}),
we assume an isothermal plasma, $p=T\rho$, in case i) (i.e.~mass conservation)
and $p=0$ in all other cases.

All quantities are normalized to a typical value of the magnetic field
strength, $B_0$, a length scale, $L_0$, and the Alfv\'en velocity,
$v_A=B_0/\sqrt{\mu_0 \rho_0}$ with a typical value for the mass
density, $\rho_0$.  A normalization time scale follows as $t_0 =L_0/v_A$, 
and the relationship of these scales with the experimental setup are
described below.

\subsection{Initial conditions}

At $t=0$, the plasma is assumed to be at rest, $\vec{v}=0$, with a
prescribed density distribution according to one of the specific
density models.  Hence, we make no attempt to describe the
ionization stage of the experiment, but rather start with a
configuration that contains highly conducting plasma in the entire
domain.

A cartesian coordinate system is used in which the plasma chamber wall
that contains the two electrodes lies in the plane $z=0$, and the
electrodes themselves are located at $\pm R \vec{e}_y$.

The initial magnetic field consists of two parts: The contribution
from the horse shoe magnet in the experiment is modeled by the field
$\vB_d(\vx)$ created by two virtual magnetic dipoles located outside
the domain at positions $\vec{x}_\pm = \pm R \vec{e}_y - R\vec{e}_z$, i.e. 
``below'' the electrodes. They are assumed to
carry dipole momenta $\vec{m}_\pm = \pm m\vec{e}_z$ with $m$ chosen
such that the field at the electrode
positions is normalized, $|\vec{B}_d(\pm R \vec{e}_y)| = 1$.  It follows
that
\begin{equation}
\vB_d(\vx) = \sum_{i=\pm} \frac{3\left(\vm_i\cdot\left(\vx-\vx_i\right)
  \left(\vx-\vx_i\right)\right)-\vm_i\left|\vx-\vx_i\right|^2}{\left|\vx - \vx_i\right|^5}.
\end{equation}

A second initial magnetic field component, $\vB_c(\vx)$, mimics the
field related to the plasma current in the early stage of the discharge.  A
quantitative model for $\vB_c$ is constructed by means of a vector
potential $\vec{A}_c(\vx)$ with $\vB_c(\vx) =
\nabla\times\vec{A}_c(\vx)$, and the choice of $\vec{A}_c(\vx)$ is
motivated by the assumptions that i) the initial current is
approximately localized around the half-circle in the plane $x=0$
which connects the electrodes,
and
ii) the direction of $\vec{A}_c$ coincides with the direction
of $\vB_d$ because the ring current will roughly follow those field lines. With
$\vec{e}_d(\vx) := \frac{\vB_d(\vx)}{|\vB_d(\vx)|}$ given from the dipole field model above,
we choose
\begin{equation}
\vA_c(\vx) = A_0 \; e^{- (|\vx|-R)^2 / \delta^2 } \; e^{-x^2/\delta^2} \; \vec{e}_d(\vx)
\label{eqn:vector_potential}
\end{equation}
The exponentials localize the magnetic structure on a scale of
$\delta =0.625$ around the ring of radius $R$ in the
$y$-$z$-plane.
The amplitude $A_0$ is determined from the condition that the maximum of $|\vB_c|$ in the
bottom plane $z=0$ is $\max_{z=0}( |\vB_c|)=3$, which roughly
corresponds to the ratio of the current-induced magnetic field to the
horse-shoe magnetic field magnitude taken at the position of the electrodes
as estimated for the experimental setup (see below).

At this point, we would like to stress that the resulting initial
magnetic field, $\vec{B}(\vx, t=0) = \vB_d(\vx) +
\nabla\times\vec{A}_c(\vx)$, is not force-free.  Even with a (small)
thermal pressure term added, there is no force equilibrium at the
start of the simulation, which is in accordance with the experiment.  The aim
here is, similar to the experiment, to investigate dynamical
properties of the resulting evolution.

\subsection{Density models}

In the next section, we will describe results obtained from four
simulation runs, all of which use the same initial magnetic field as
described above, but differ in the treatment of the mass transport,
i.e. in the specific realization of Eq. (\ref{eqn:MHD-cont}):

\bigskip

{\bf Case 1 - Mass conservation:}
Here, Eq. (\ref{eqn:MHD-cont}) is used as a continuity equation with $S=0$, describing mass
conservation. More specifically, we use a homogeneous mass density $\rho=1$ as initial condition and
specify the temperature as $T=1$. This model can be seen as the simplest choice possible that accounts
for a consistent mass transport during the dynamical evolution of the system. 
Realizing that $| \vB | \approx 3$ in the current filament, the resulting local plasma-$\beta$
is $2  / | \vB |^2 \approx 0.2 \ll 1$.

\bigskip

{\bf Case 2 - Fixed density:} We keep $\rho=1$ fixed throughout the
simulations, i.e. Eq. (\ref{eqn:MHD-cont}) is abandoned. As a
consequence, sound waves are eliminated from the dynamics. This model is used in order to get
an indication of the influence of the mass transport and pressure included in the previous case
on the evolution.

\bigskip

{\bf Case 3 - Fixed Alfv\'en velocity:} In this case, the mass density
is continuously adjusted such that the Alfv\'en velocity is constant
throughout the domain, i.e. $\rho(\vx, t) \propto |\vB(\vx, t)|^2$.
The interest for this study stems from the fact that the evolution might be
treated as quasi-static \citep[][]{bib:bellan2000}, which means that the
Alfv\'en crossing times are small compared to the global evolution time
scale. Setting $v_A$ to a constant value results in a homogeneous
communication of Alfv\'enic disturbances.

\bigskip

{\bf Case 4 - Ionization/recombination model:}
Here, we implemented a simple model for the ionization by the electric current and chose  the source term
as
\begin{equation}
S = \Gamma_i \vj^2 + \Gamma_r (\rho - \rho_0).
\end{equation}
The coefficients are set to $\Gamma_{i}=0.5$, $\Gamma_r=5$ and
$\rho_0=1$ so that the time scales of ionization and recombination are
comparable to the Alfv\'enic and convection time scales.

\bigskip

\subsection{Numerical implementation}

Equations (\ref{eqn:MHD-cont})--(\ref{eqn:MHD-induction}), with a
specific density model, are discretized by finite-differences on a
cartesian grid and integrated as an initial value problem using a
third order Runge-Kutta scheme. The numerical box covers $[-20,
20]\times[-20, 20]\times[0, 40]$ in the $x-$, $y-$ and
$z-$direction, respectively. Boundary conditions at $z=0$, i.e. the
``bottom'' plane, are such that $\vv=0$ and
$\vec{B}_\perp$ is linearly extrapolated.
We use outflow boundary conditions, i.e. the ghost cell values are set to the
first cell values within the domain.  However, the simulated
current filaments are well separated from those boundaries so that
these conditions have no significant influence on the results.

\begin{figure}[ht!]
  \begin{center}
    \includegraphics[width=1.0\columnwidth]{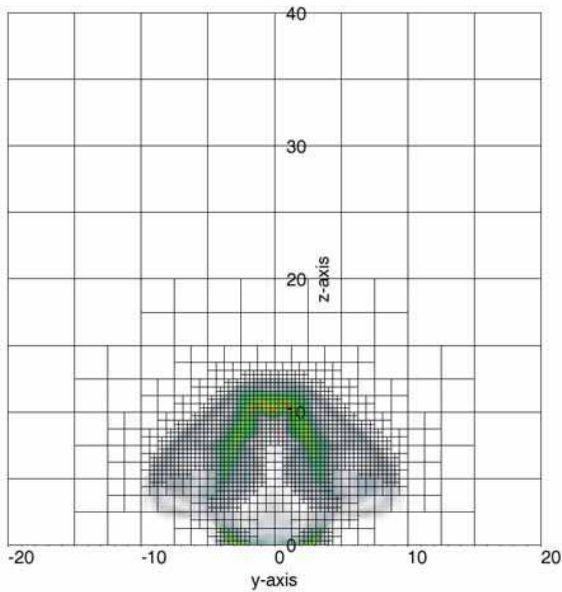}
  \end{center}
  \caption{\label{fig:init-grid}
The full computational domain with the block distribution and 
the line-of-sight integrated electric current density,
viewed parallel to the $x$-axis.
}
\end{figure}

In order to obtain sufficient spatial resolution of the current
filament dynamics, we carried out mesh-adaptive computations with local
refinement using the simulation code ``racoon'' \citep[][]{bib:racoon}. 
Grid blocks of $16^3$ cells each are recursively refined
up to a local resolution equivalent to $1024^3$, using a refinement
criterion that compares the electric current density $|\vec{j}|$ with
a critical value $j_{crit}$ that is determined from the existing
local resolution.  A typical grid layout is depicted in
Fig. \ref{fig:init-grid}.

\subsection{Scaling of experimental parameters}
The model described above involves a number of normalization values
for the MHD quantities.  In order to allow a direct comparison with the
experiments of Bellan et al. and
``FlareLab,'' we relate these values to
physical parameters given in \citep[][]{bib:bellan2000} and estimates from the ``FlareLab'' experiment
(not yet published).

With $L_0 = 16 \; {\rm mm}$, the electrode spacing is $5 L_0 = 8 {\rm cm}$
which corresponds to the experimental value, and the entire domain
covered in the computation is $64\times 64\times 64 {\rm cm}$ which
roughly corresponds to the dimensions of the plasma vessel.
Taking $B_0=0.3 \; {\rm T}$, which matches the horse shoe magnetic field at the electrodes in the
``FlareLab'' 
experiment, and assuming a hydrogen plasma of $n_0\approx 10^{20} \; {\rm m}^{-3}$,
results in an Alfv\'en velocity of $v_A=6\cdot 10^{5} \; {\rm m/s}$ and an Alfv\'en 
crossing time of $t_0 \approx 2.6\cdot 10^{-8} \; {\rm s}$. This value is small compared to
the macroscopic evolution time scale of $\approx 1{\rm \mu s}$ reported for the experiment,
but two things should be noted here: 
First, from experiment the detailed density distribution
can only be estimated
as the experiments still lack suitable diagnostics to measure this value. More importantly,
the magnetic field strength drops off drastically with increasing
distance from the magnet poles/current filament, 
and the filament length increases in time from its initial value of $\approx 10 \; L_0$.
Hence, the true Alfv\'enic travel time
along the filament from one electrode to the other will be longer than $t_0$
and might reach the overall evolution time scale.

As for the magnitude of the initial magnetic field created by the
plasma current, $\vec{B}_c$, we assume the entire discharge current of
$I \approx 50 \; {\rm kA}$ to be located in a channel of radius
$\delta = 0.625 L_0 = 10{\rm mm}$. By Ampère's law, this will create an azimuthal
magnetic field of $B_\varphi = \mu_0 I / 2 \pi \delta = 1 \; T \approx 3.3 \; B_0$ 
on the channel surface. Hence, we chose the magnitude of $A_c$
in Eq. (\ref{eqn:vector_potential}) such that $|\vec{B}_c| = 3$ close to the electrode.

The viscosity in (\ref{eqn:MHD-motion}) is included for numerical
stability of the simulations. It was necessary to use a normalized
value of $\nu=0.01$ in cases 1 and 2 and an even larger value of
$\nu=0.05$ in cases 3 and 4.
These values might correspond to a significantly higher viscosity
than the actual physical one of the experiment. However, not only is
the latter largely unknown by value, but it is also questionable if
the viscous term in the momentum equation provides a realistic description
at all.
Hence, we treat the viscosity as a purely numerical parameter necessary for stabilizing the
computations.

\section{Simulation Results}

The observations in the experiment are made with an optical CCD camera. 
It is obviously difficult, if not impossible, to unambiguously relate
the luminosities of the images to plasma quantities, but we see it as
a reasonably assumption that there exists a close correlation between
the electric current density and the luminosity.
Therefore, the simulation results presented here focus mainly on the
current density evolution, and volume renderings of $|\vj |$ are
shown in the plots discussed below for comparison with the images
taken by the laboratory camera.
Further, a view perpendicular to the current arc is chosen, again to
fit the camera's perspective in the ``FlareLab'' set-up.

\begin{center}
\begin{figure*}
\includegraphics[width=0.9\columnwidth]{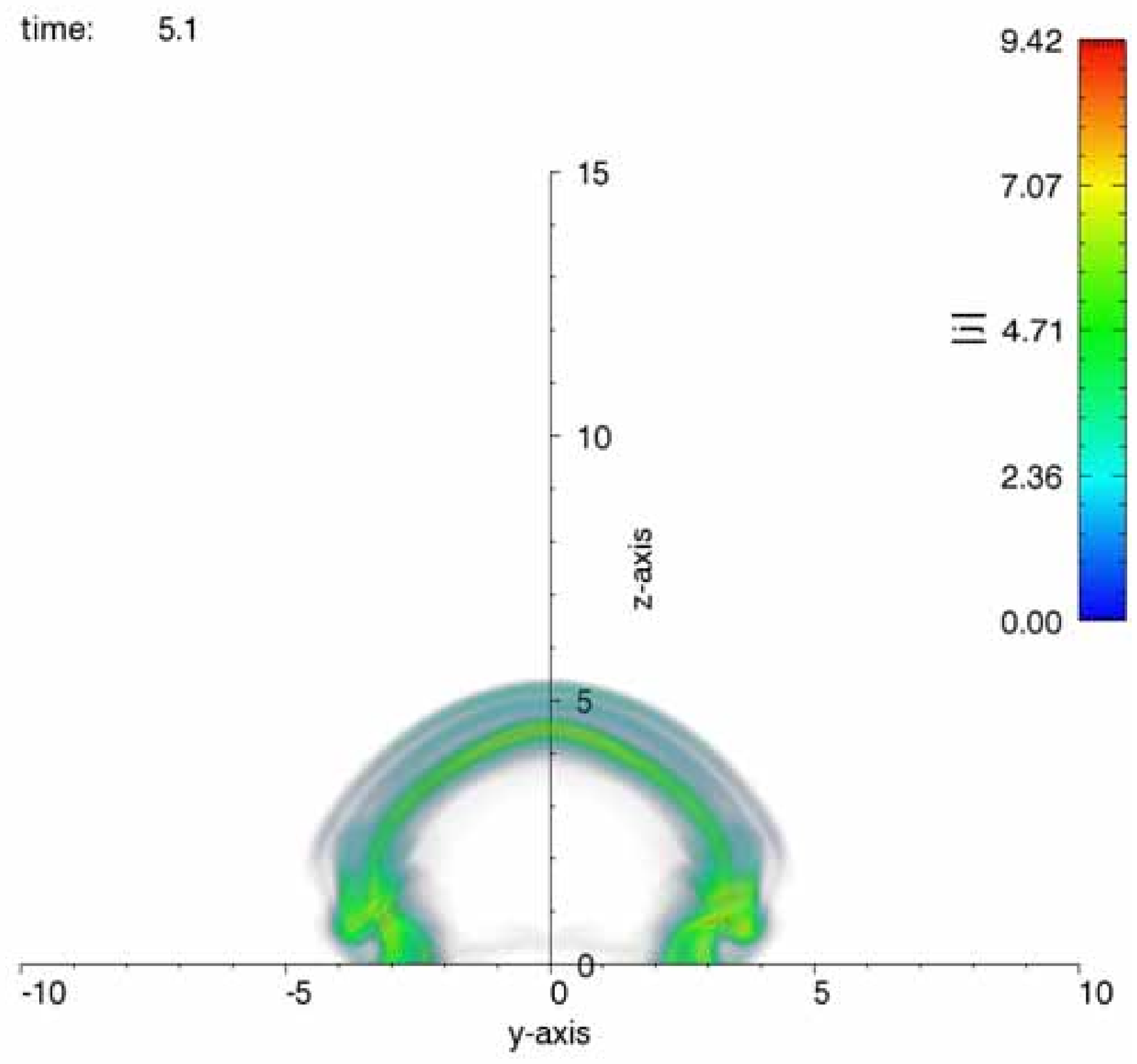}
\includegraphics[width=0.9\columnwidth]{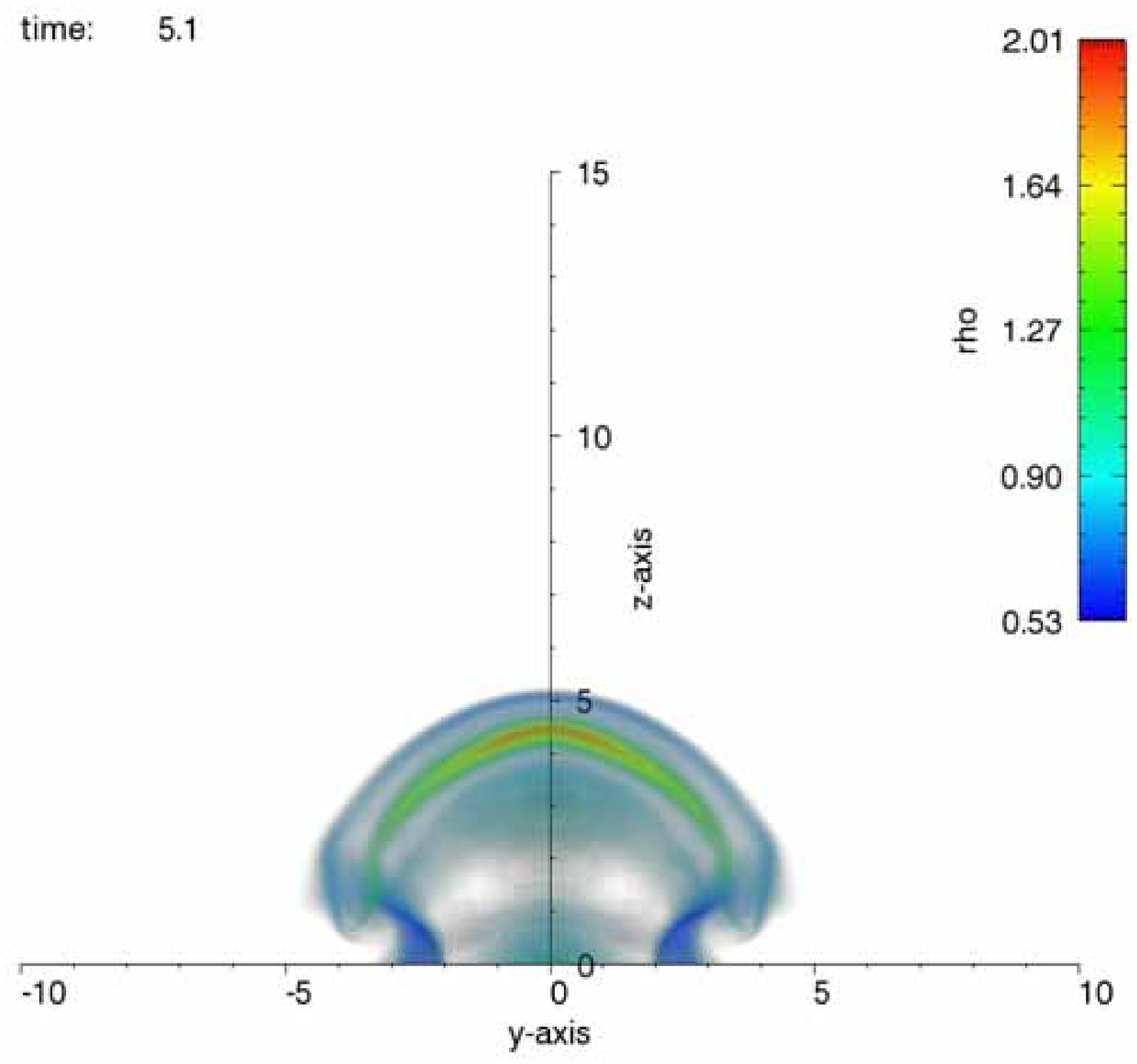}\\
\includegraphics[width=0.9\columnwidth]{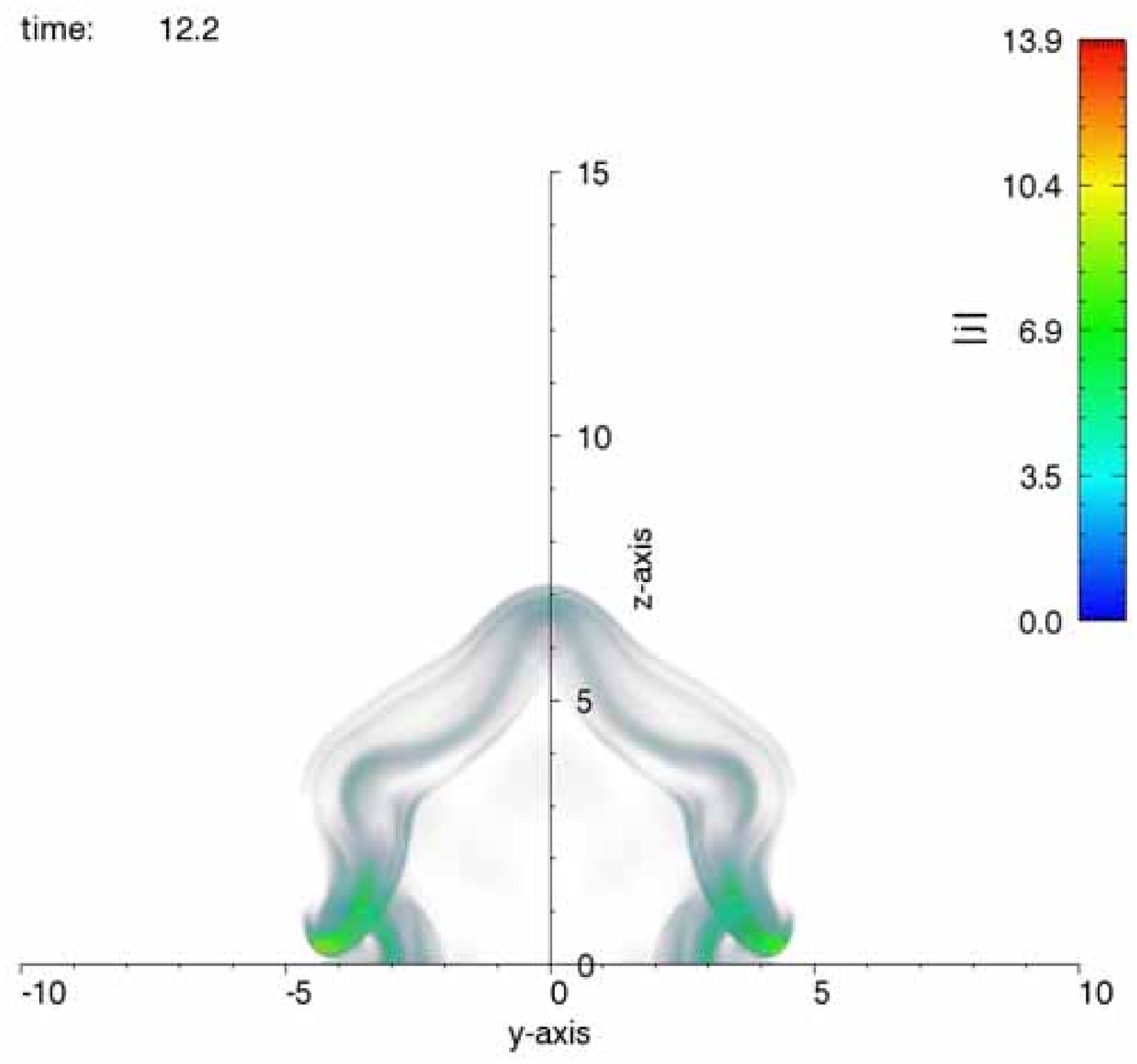}
\includegraphics[width=0.9\columnwidth]{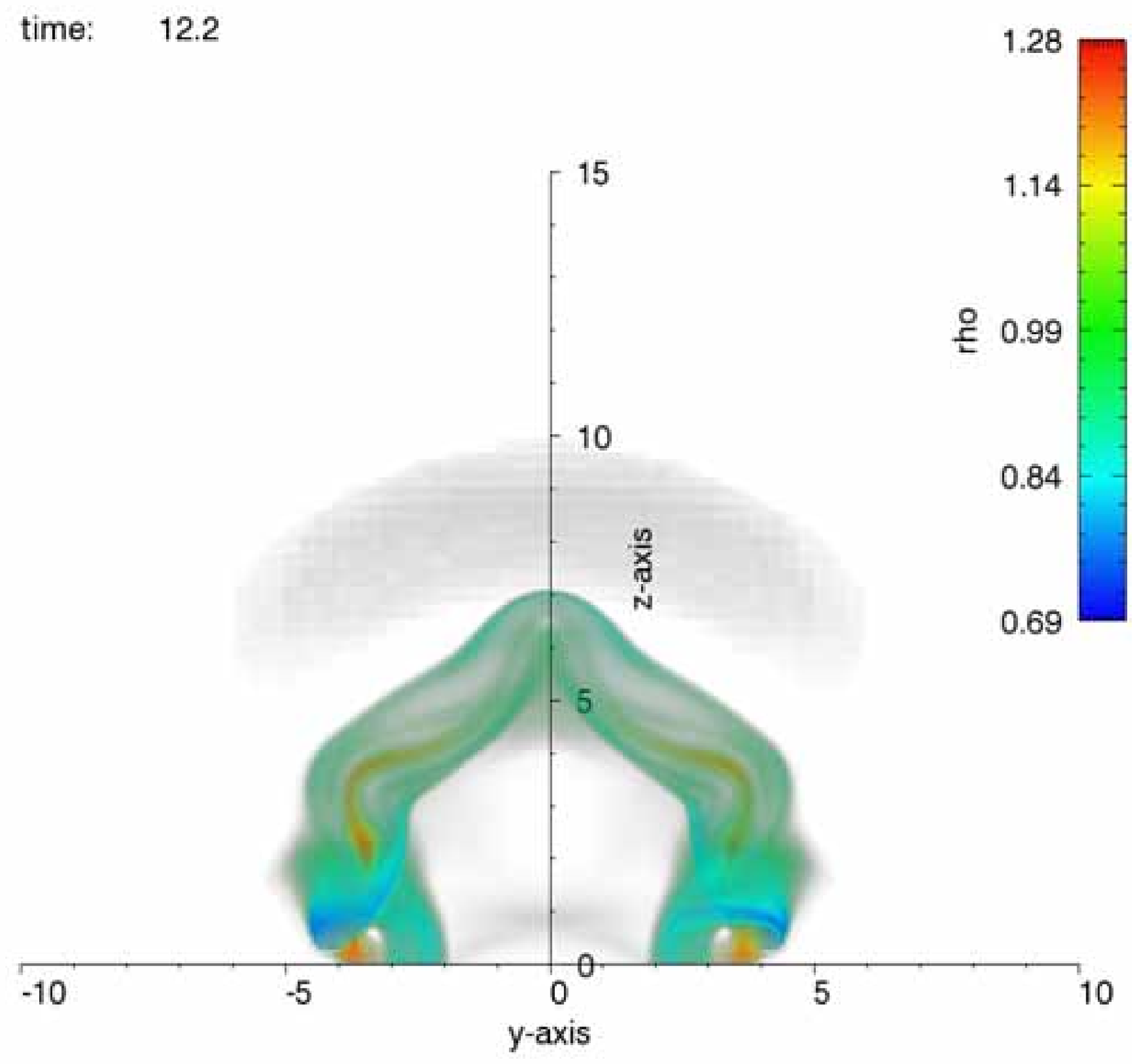}\\
\includegraphics[width=0.9\columnwidth]{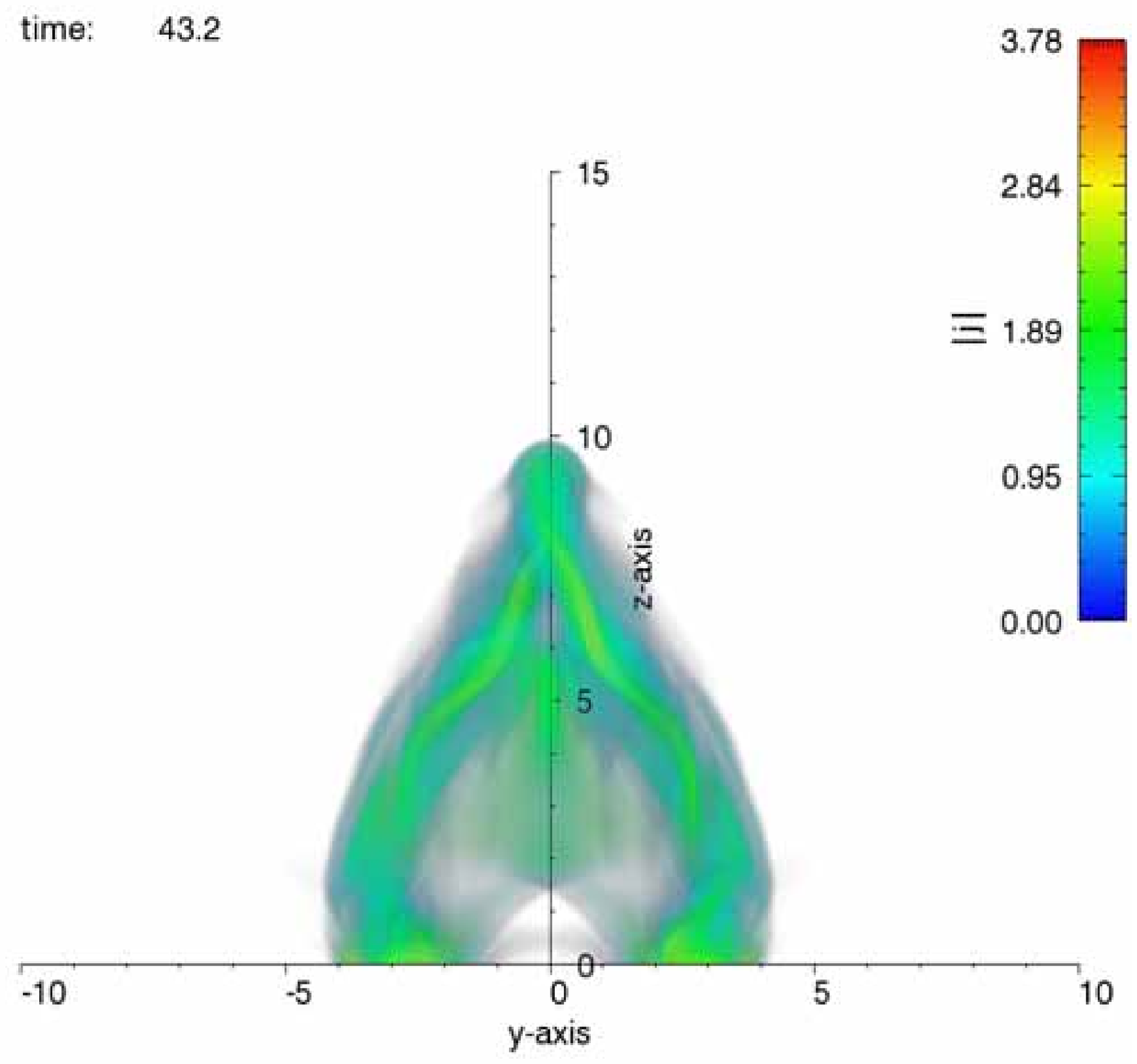}
\includegraphics[width=0.9\columnwidth]{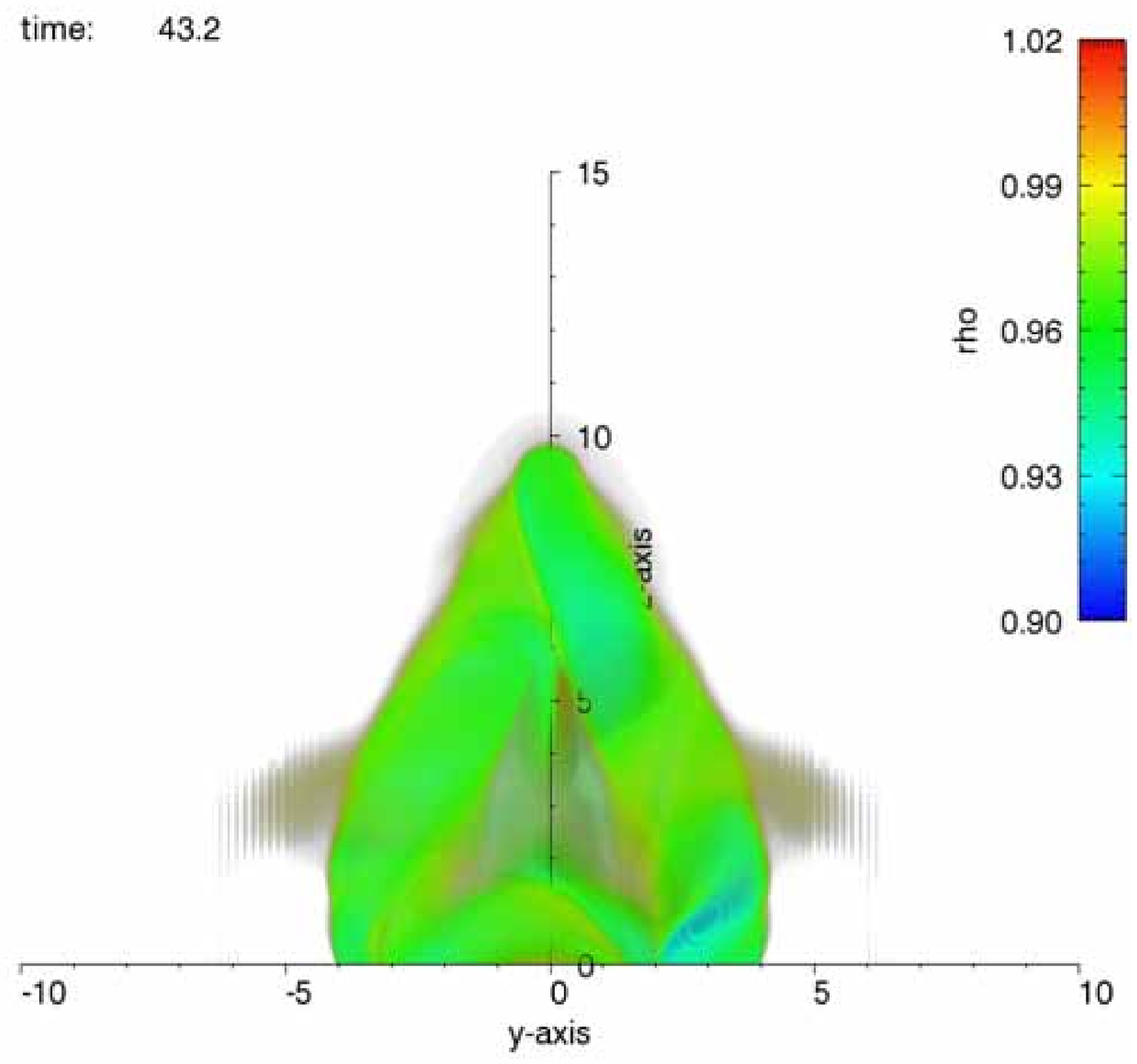}\\
\caption{
\label{fig:case1}
Electric current density $|\vj|$ (left) and mass density $\rho$ (right)
of simulation run 1
at times $t=5.1$ (top), $t=12.3$ (center) and $t=43.2$ (bottom).
}
\end{figure*}
\end{center}

Fig. \ref{fig:case1} shows the spatial distributions of $|\vj|$
and $\rho$ at three different times of simulation case 1,
i.e. the run implementing mass conservation.
Essential dynamical features can be observed on different time scales:
Within the first Alfvén time, the current arc pinches to
approximately half of its initial diameter. 
This effect can be determined from the plots taken at $t=5.1$: The
current is more localized than in the initial configuration, and
the mass density is enhanced in the arc center as a consequence of the
compressional plasma motion.
Accordingly, the density is reduced around the arc with minimum values
of $\approx 0.53$.  This deviation of mass density distribution from
the initial homogeneous value corresponds to the formation of pressure
gradients due to the assumption of isothermal plasma with normalized temperature $T=1$.
It should be noted that the pinching is less pronounced close to the
electrodes where the dipole magnetic field component is comparable in
magnitude to the non-potential $\vec{B}_c$.
Apart from this pinching, the arc as a whole has undergone an
expansion from its original major radius $R=2.5$: The apex has risen
to $z\approx 4.5$ and the curvature in the upper part of the arc is
reduced compared to the original half-circle shape.  Also, close to
the electrodes, a more pronounced deformation of the current channel
into helical shape is observed.
Both, the expansion and the foot point deformation occur on the scale
of several Alfvén times, i.e. slightly slower than the pinching.
In the following evolution, the current structure and the mass density
perturbation continue to expand (resp. times $t=12.2$ and $43.3$),
causing the loop apex to rise to about $z=10$, i.e. four times its
initial height.
In the course of this expansion, the tendency of helical deformation
manifests itself on the entire structure and grows in amplitude.
Eventually, the upper part of the loop is entirely rotated out of the
plane $x=0$ and intersects that plane at right angle with its apex
(time $43.2$).
Later, the expansion slows down and finally stalls shortly
after the last frame shown in Fig. \ref{fig:case1}.
After that, the structure actually starts to shrink slightly but finally comes
to rest in a configuration close to the last one shown.

The qualitative interpretation of this sequence in terms of ideal MHD
appears to be straight-forward: Recalling that the initial
configuration is not in force balance, the current will pinch towards
a force balance perpendicular to the center of the current tube.
With the original diameter of $2\delta= 1.25$ and an amplitude of the
current-induced field of $|\vec{B}_c|=3$ and density $\rho\approx 1$, 
this process occurs on a time scale of $\approx 0.1$.
The expansion of the arc can be interpreted as the response to the
well-known ``hoop force'' caused by the arc curvature.  As the arc
follows this force, the curvature and the corresponding current
density are reduced and the expansion slows down.  In the vicinity of
the foot points, the line-tying condition and the zero velocity
boundary condition prevent the arc from following the expansion in the
horizontal directions, and hence the arc looses its circular shape
with its upper part merely rising upwards.
Finally, we interpret the formation of helices at the foot points as
kink modes and the fact that these modes develop fastest at the
electrodes as a consequence of the different local Alvf\'en velocities:
While the magnetic field is dominated by $\vec{B}_c$ on the entire
arc, the mass density is considerably reduced to values around $0.5$
close to the electrodes in the early stage dynamics (cf. $\rho$ at
$t=5.1$ in Fig. \ref{fig:case1}). Hence, the Alfv\'enic kink dynamics
will be twice as fast as in the apex, where $\rho\approx 2$ as a
consequence of the early pinching.  This estimate is consistent with
the fact that kinks become apparent at $t\approx 10$ at the apex,
while they form already around $t\approx 5$ at the electrodes.
At later stages, the entire structure relaxes into an approximately
force-free state in which the internal twist of the magnetic field has
been converted into a large-scale writhe.

Using the results of case 1 from above as a reference, we describe in
the following the significant differences observed in the remaining simulation runs.
\begin{center}
\begin{figure*}
\includegraphics[width=\columnwidth]{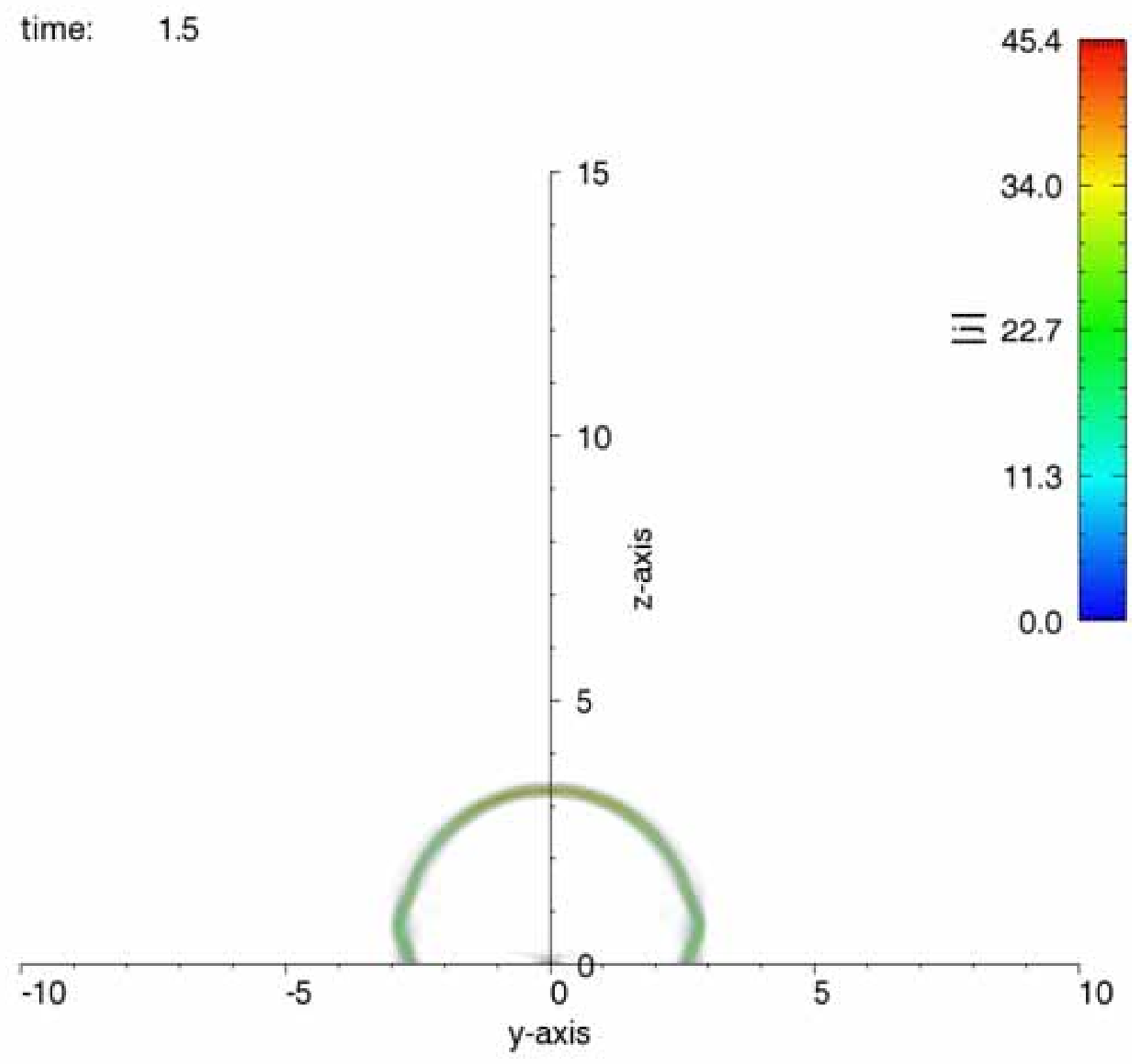}
\includegraphics[width=\columnwidth]{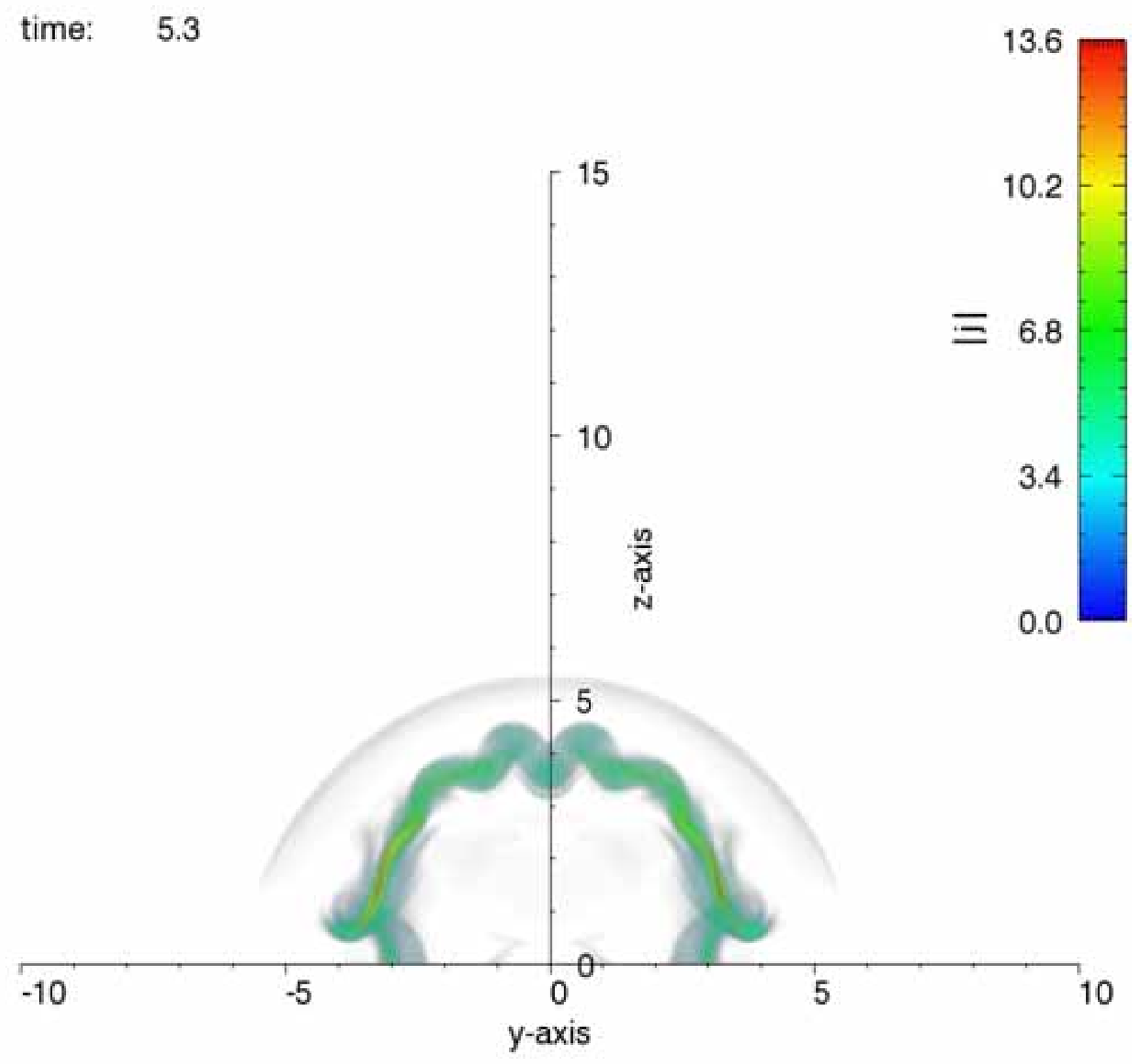}
\includegraphics[width=\columnwidth]{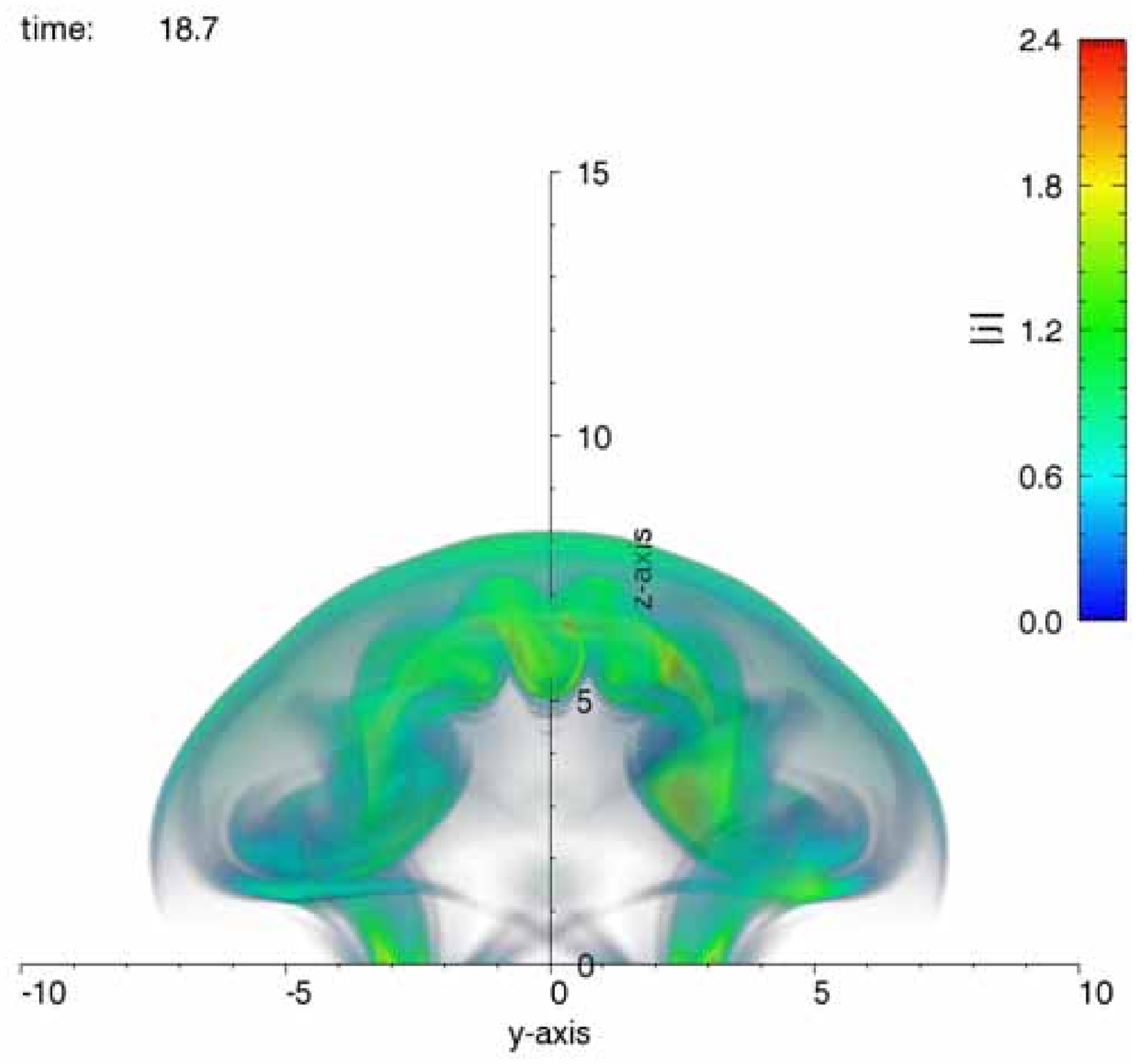}
\includegraphics[width=\columnwidth]{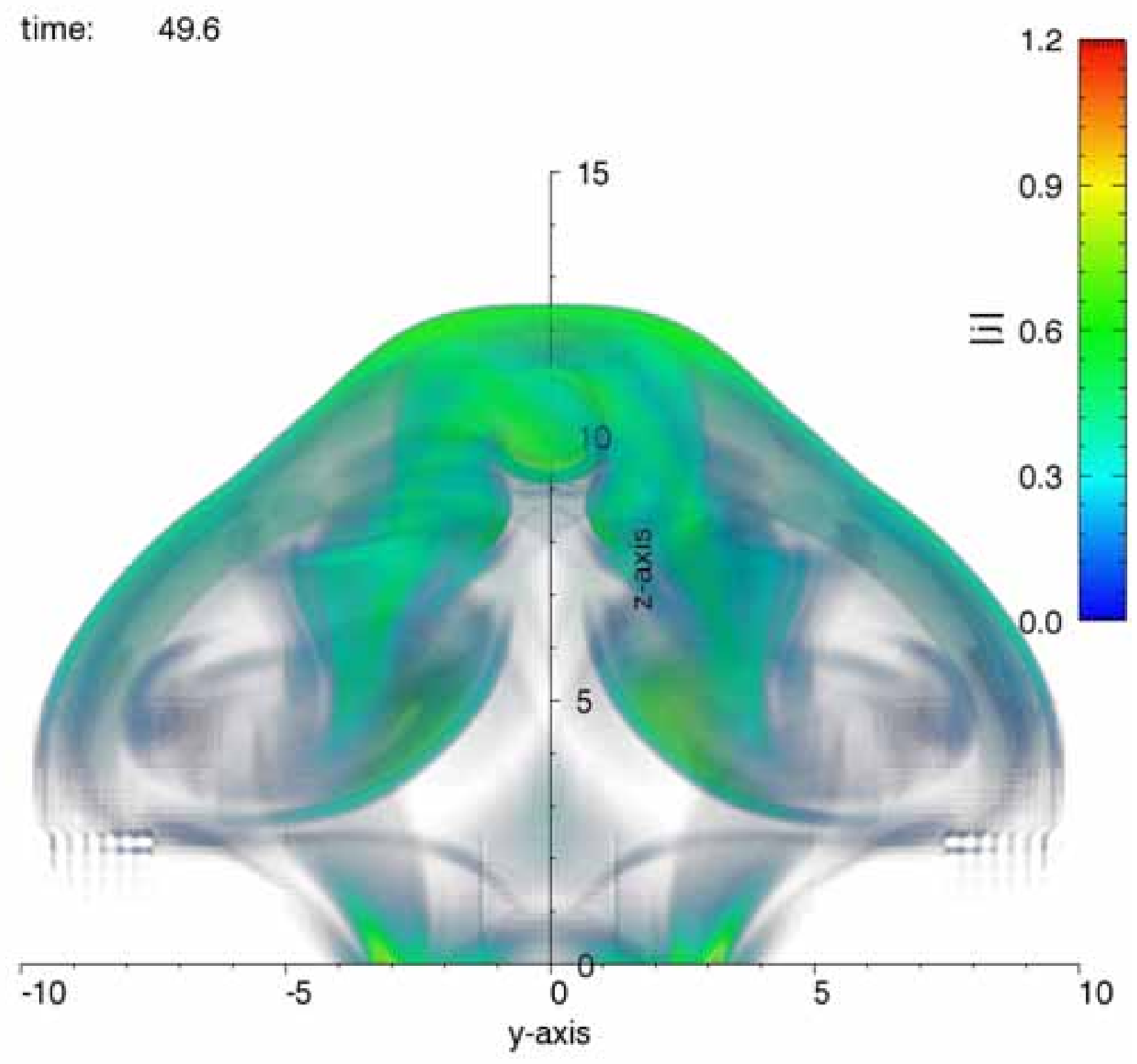}
\caption{
\label{fig:case2}
Electric current density $|\vj|$ of simulation run 2
at times $t=1.5$, $t=5.1$, $t=18.7$ and $t=49.6$.
}
\end{figure*}
\end{center}

Fig. \ref{fig:case2} displays the current structure of case 2 (``fixed density'') at different times.  
While the overall evolution is similar to case 1, some consequences of the 
constant mass density, and hence the absence of pressure forces, are readily visible.
First of all, the maximum current density during the pinching phase is
significantly larger than in case 1 ($\approx 45$ compared to $\approx
10$ at time $t\approx 1.5$), apparently because of the absence of
restoring thermal pressure forces in case 2.
Another consequence of the homogeneous mass density is that the kink
formation in the apex region occurs already at $t\approx 5$, at the
same time as it is observed close to the foot points.
This homogeneity along the arc slightly alters the overall structure
formation: While case 1 displayed an evolution from the foot points to
the apex and from short wavelengths to larger scales, the development
in case 2 is almost uniform along the arc. Small scale kinks occur
early and grow in amplitude until the perturbation of the arc becomes
almost stationary in shape, and only the slow expansion of the
structure can be seen in the simulation. In particular, the pronounced
rotation of the structure as a whole as seen in case 1 is absent
here. 
Rather, the late stage can still be interpreted as a large current arc
lying in the $x=0$-plane and continuing to expand.

\begin{center}
\begin{figure*}
\includegraphics[width=\columnwidth]{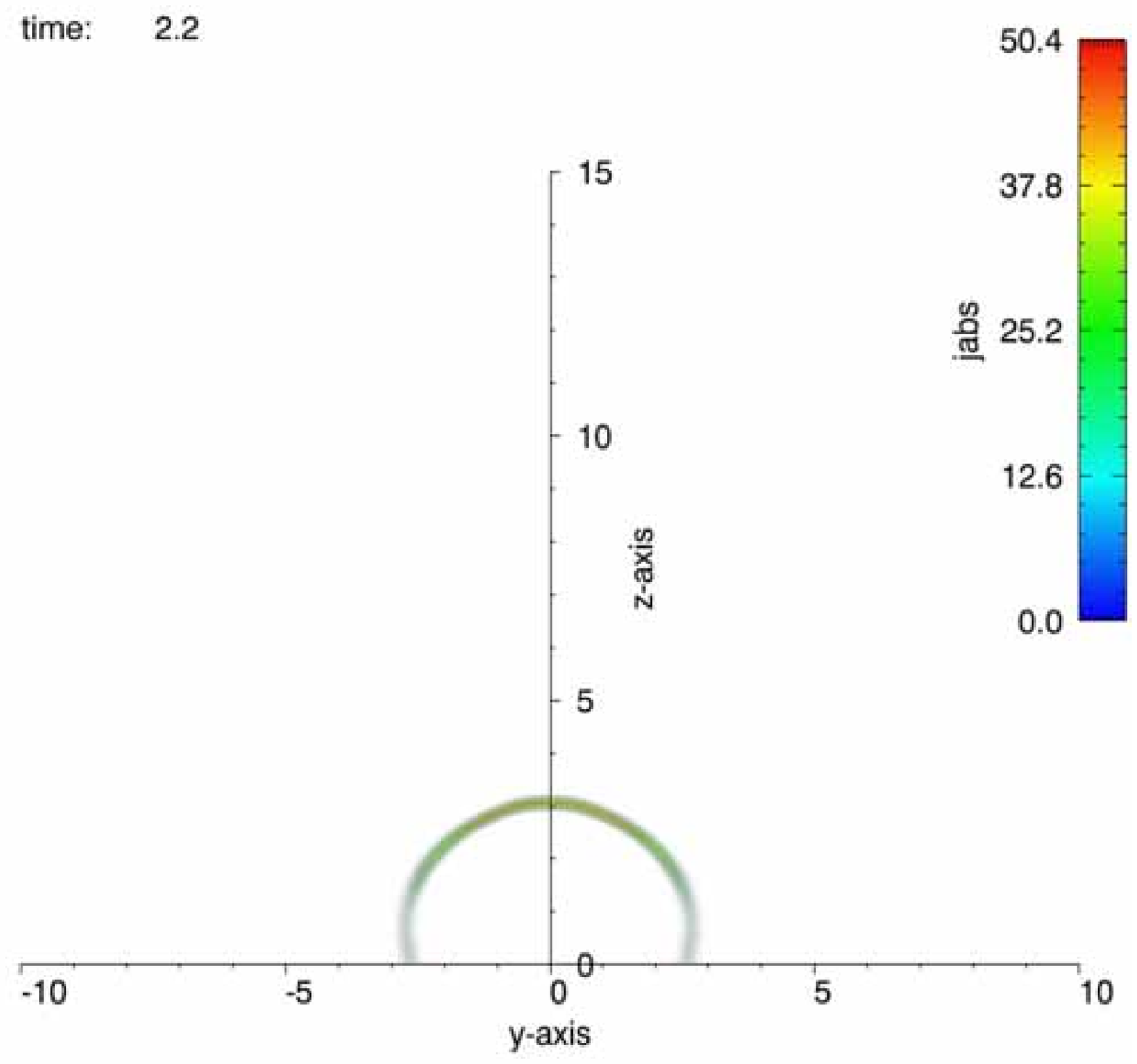}
\includegraphics[width=\columnwidth]{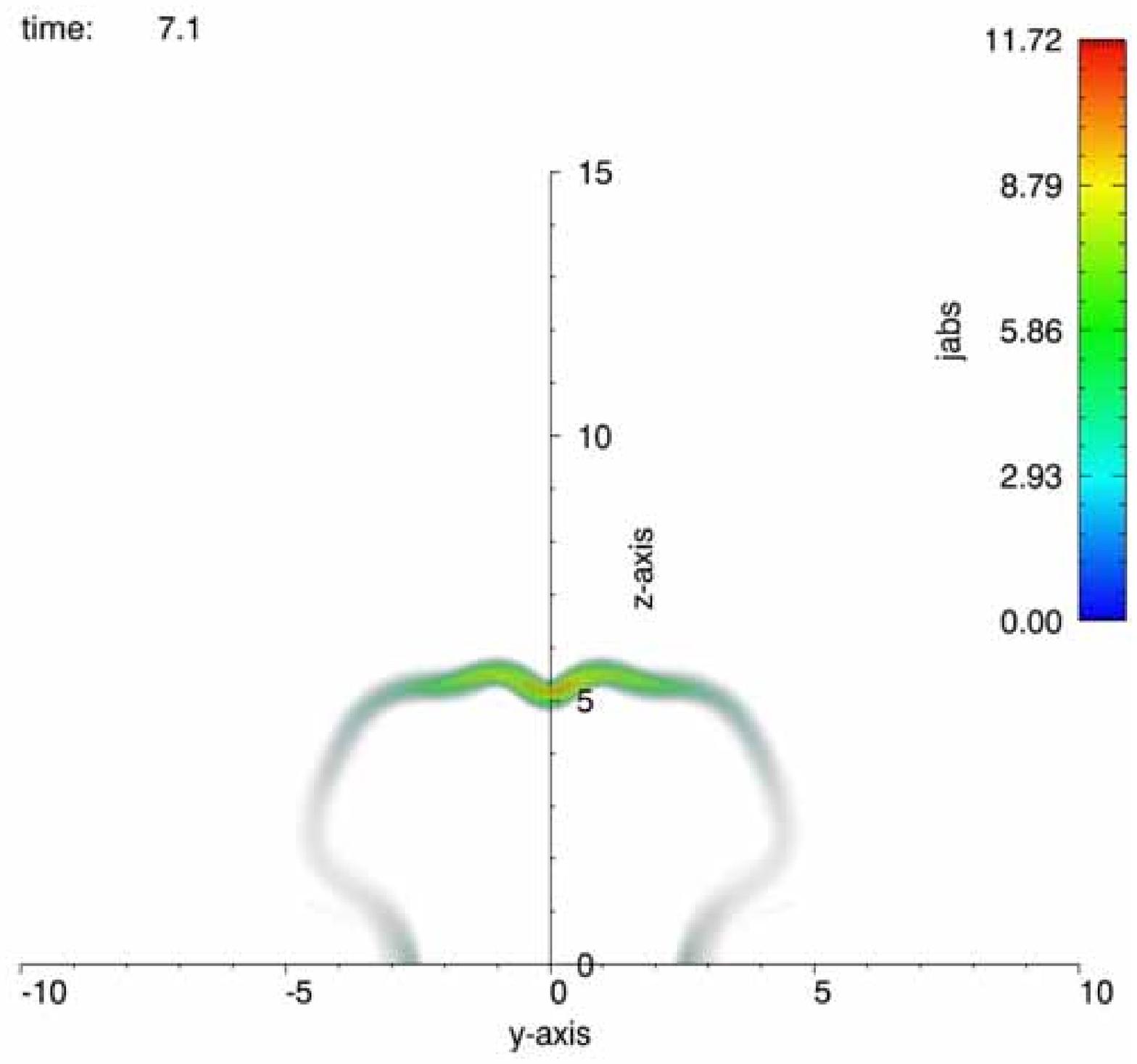}
\includegraphics[width=\columnwidth]{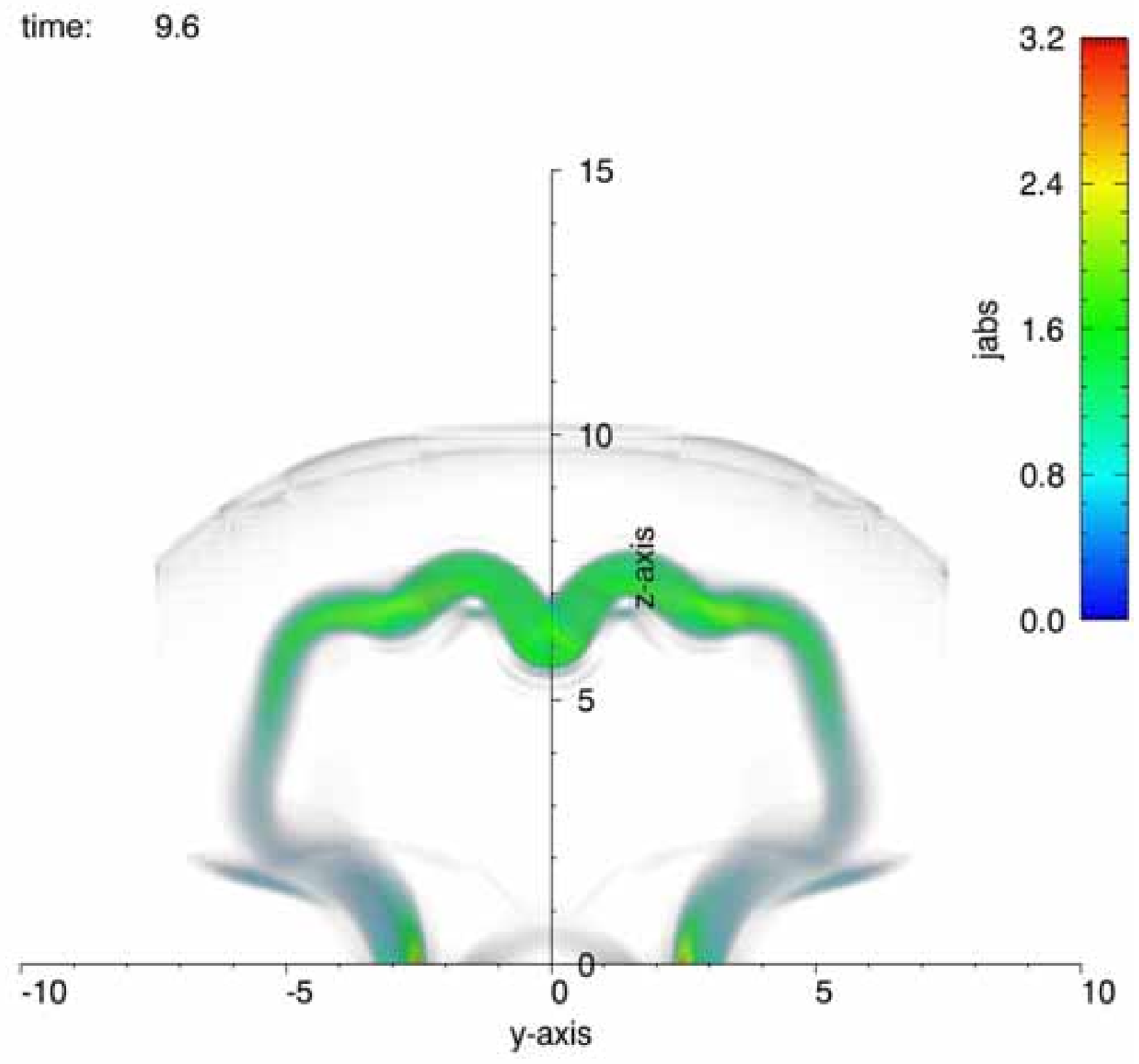}
\caption{
\label{fig:case3}
Electric current density $|\vj|$ of simulation run 3 at times $t=2.2, 7.2$ and $9.6$, respectively.
}
\end{figure*}
\end{center}

Fig. \ref{fig:case3} shows that the evolution of the current structure
in case 3 (``fixed Alfv\'en velocity'') resembles closely the results of case 2. The upper part of
the arc flattens and forms kinks, however, the formation of kinks is
constrained to the upper part of the arc and the ``legs'' remain
almost unperturbed.

\begin{center}
\begin{figure*}
\includegraphics[width=\columnwidth]{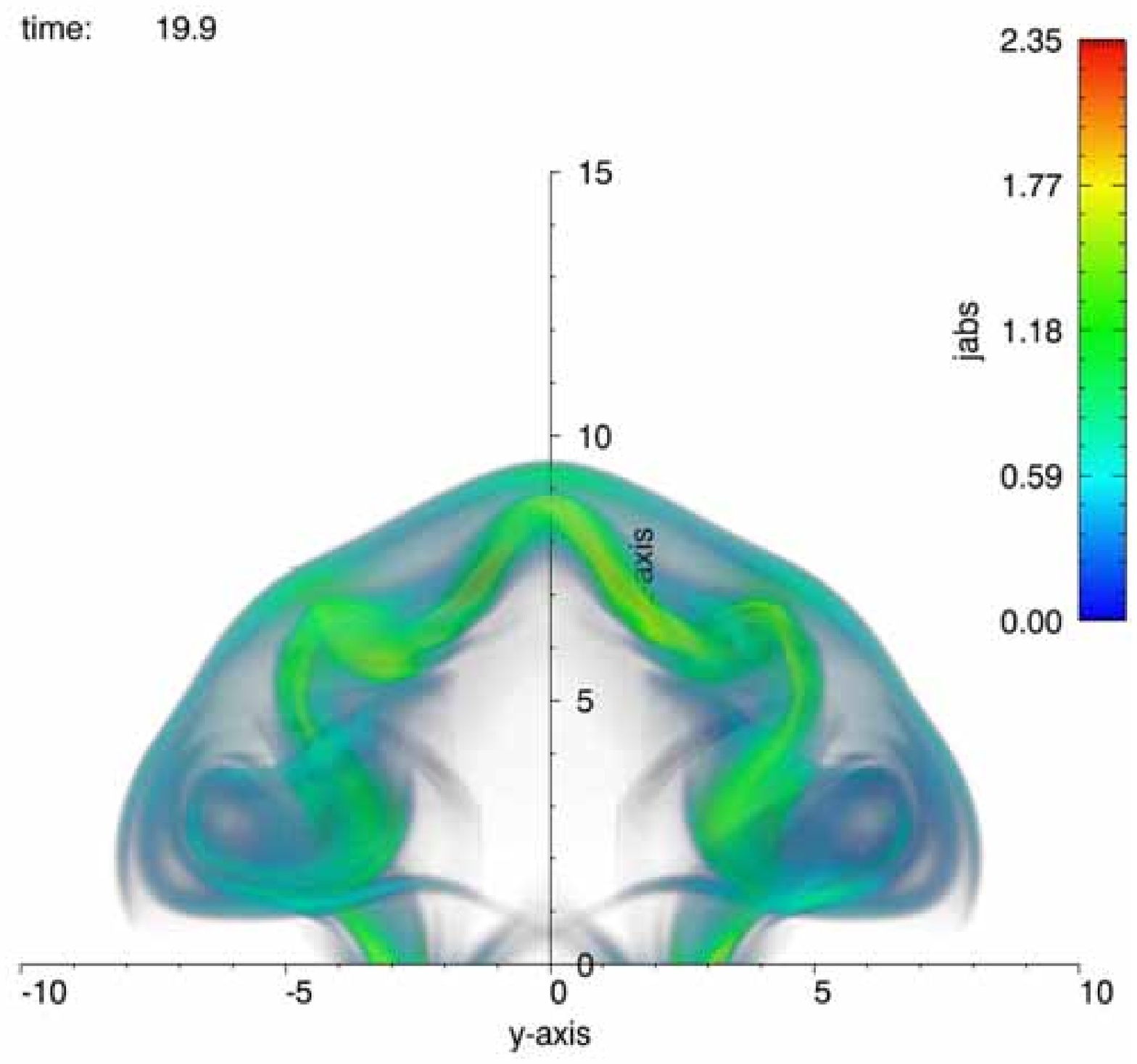}
\includegraphics[width=\columnwidth]{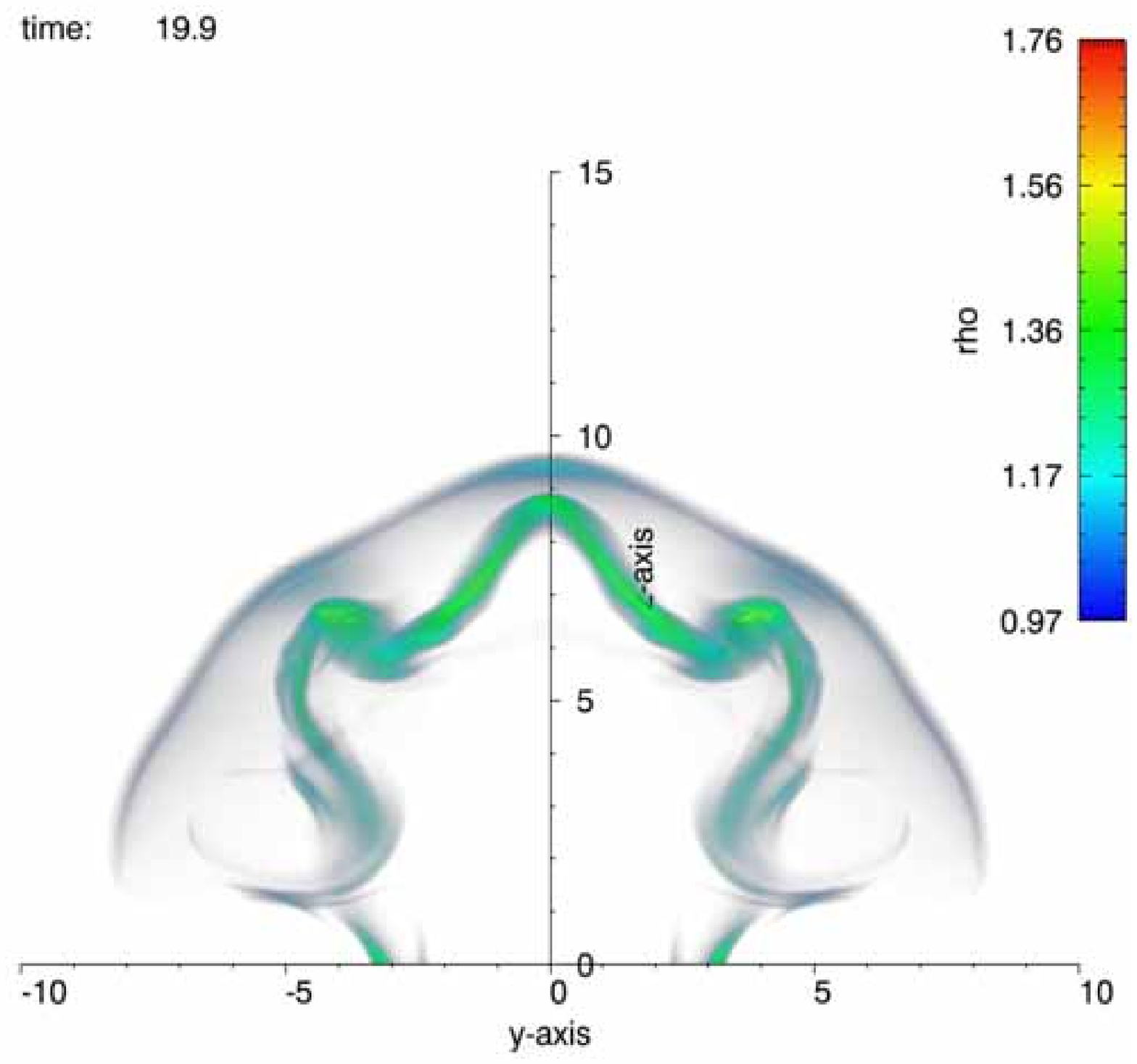}
\caption{
\label{fig:case4}
Electric current density $|\vj|$ (left) and mass density (right) of simulation run 4, time $t=19.9$.
}
\end{figure*}
\end{center}
Finally, the inclusion of a simple recombination/ionization model in
case 4 again leads to a slight differences in the resulting
current system (Fig. \ref{fig:case4}), but the overall shape is
still comparable to the previous cases, in particular to case 2.

\subsection*{Comparison with Experiment and Discussion}

As noticed before, the experimental observations and the numerical simulations can 
only be compared on a qualitative level, not least because of the missing quantitative 
data from the experiment: 
It is not clear in which way the photographic luminosity originating
from the neutral hydrogen emissions is related to the physical plasma
parameters and fields.
Based on the assumption that the electric current is the dominant
source for ionization, and thus the plasma density and temperature
will be well correlated to the current density, we choose to compare
the laboratory images to $|\vec{j}|$ obtained from the numerical
simulation.

Further fundamental differences between the experimental regime and
the simulation model must be kept in mind: First of all, the model of
ideal MHD is an idealized approach to the physics of the plasma
discharge, in particular with respect to the degree of ionization, the
ionization process during ignition, the role of collisions and
possible kinetic or at least multi-fluid effects.
In addition, the treatment of the boundaries and the coupling of the
electrodes to the plasma is highly idealized. An other difference
between model and experiment is that the total discharge current
increases in time during the arc evolution, while the simulation boundary
conditions imply a constant, prescribed current through the lower
plane.
Concerning the time scales, we note that a direct identification of
simulation time with observed time scales is difficult because the
fundamental plasma parameters of the experiment are currently only
estimated.

\begin{figure*}[ht!]
\begin{center}

\begin{tabular}{cc}
\begin{minipage}{1.\columnwidth}
\includegraphics[width=\columnwidth]{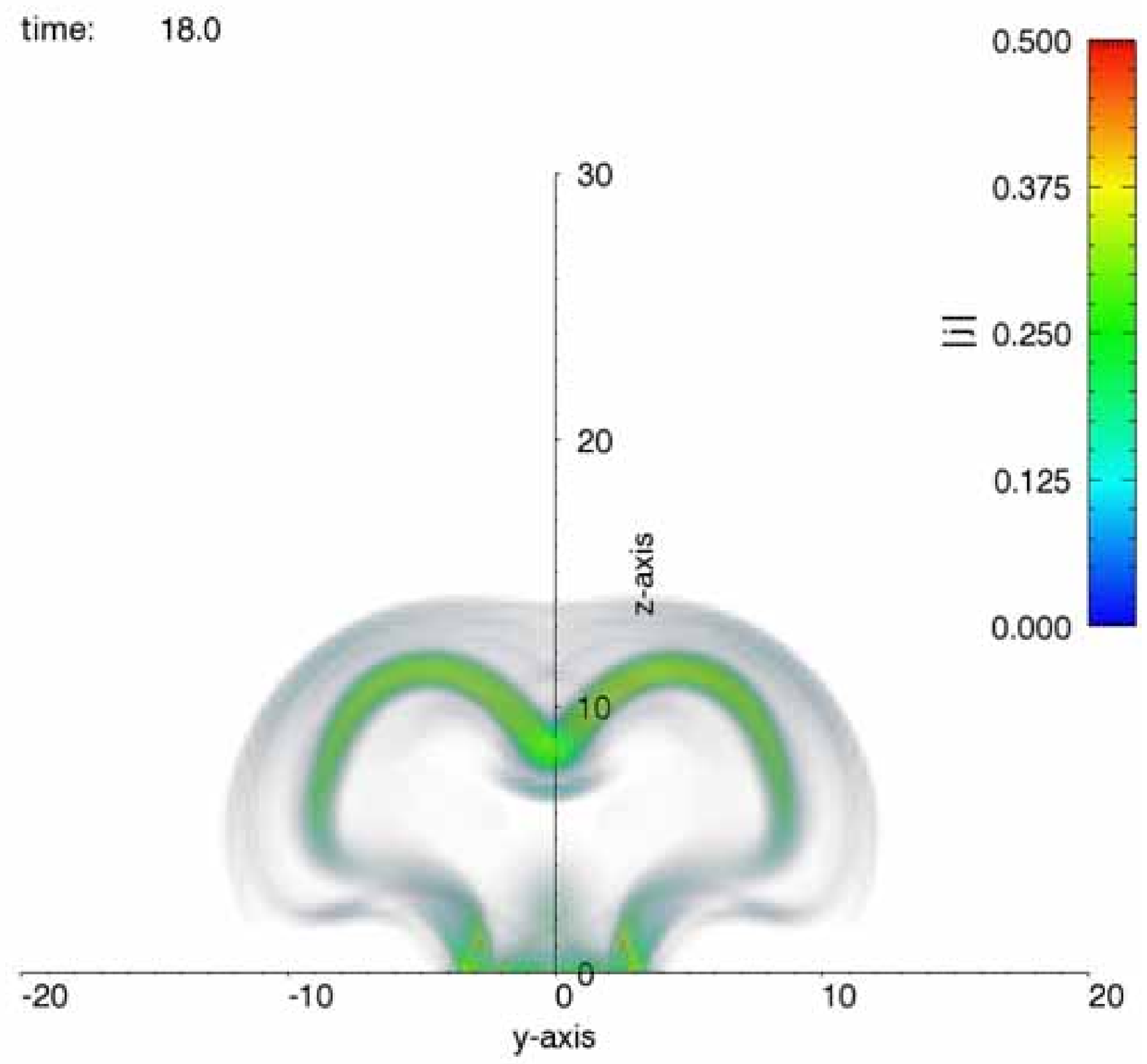}
\centerline{(a)}
\end{minipage} \\
&\\
&\\
\begin{minipage}{1.\columnwidth}
\includegraphics[width=\columnwidth]{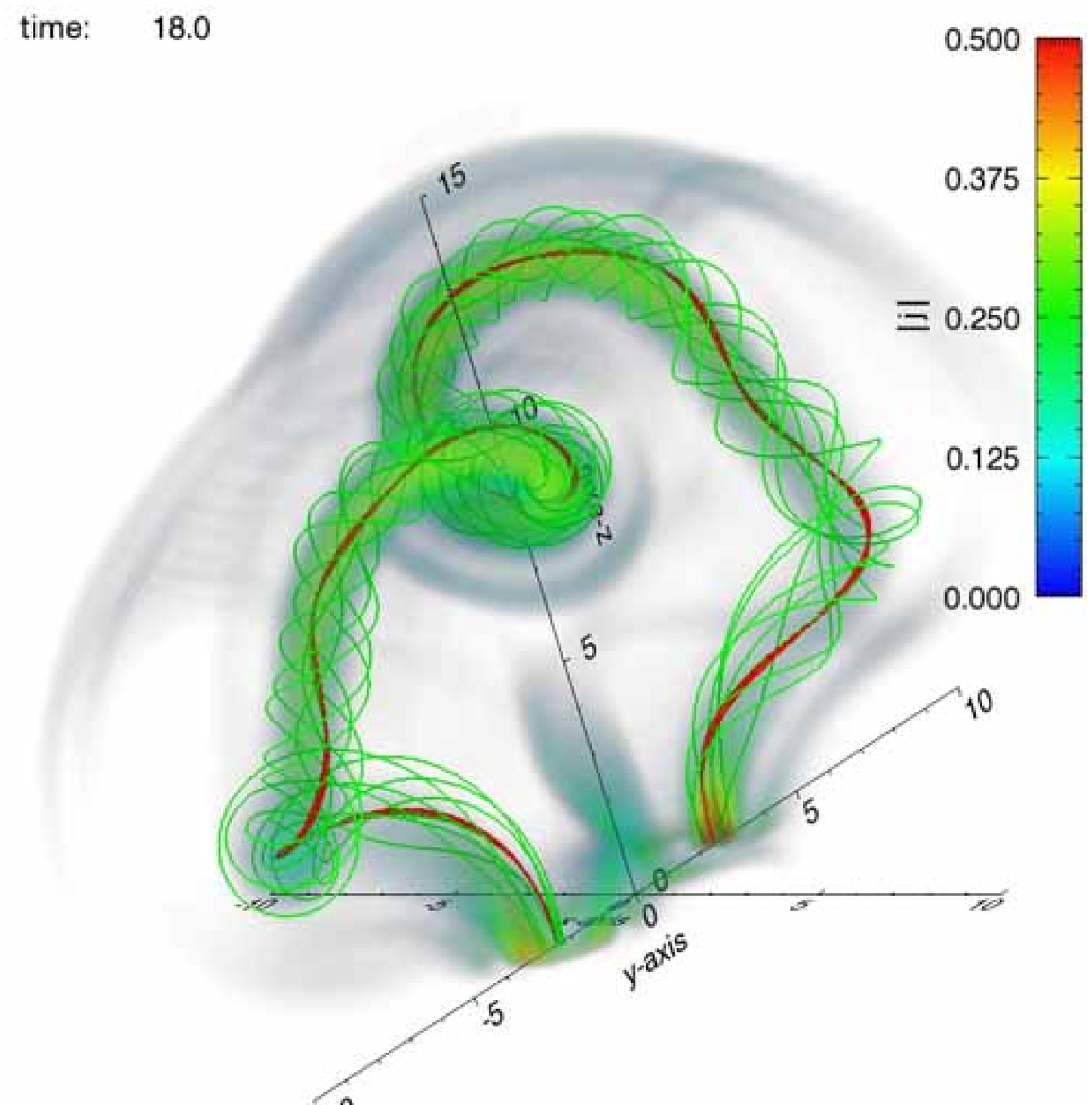}
\centerline{(b)}
\end{minipage} &\\
\end{tabular}

\caption{
a) $|\vec{j}|$ from run similar to case 3 but with $\nu=0.1$ at time $t=18.0$.
b) Perspective view of selected magnetic field lines (green), electric current lines (red)
and $|\vec{j}|$ (grey shade) from the same run.
\label{fig:comp}
}
\end{center}
\end{figure*}

Keeping these limitations in mind, the computations reproduce key
features of the laboratory results like the arc expansion and the kink
formation.
For example, by comparing Fig. \ref{fig:flarelab} with the
distributions of $|\vec{j}|$ from the four simulation runs, the
qualitative agreement becomes evident.
The moderate modifications of the density model that we employed in
the four different runs lead to slightly different details in the arc
dynamics as manifested, e.g. in the location, number and intensity of
kinks.
Asking which of these models can best reproduce the experiment, we
observe that a characteristic feature there is the fact that the
central region of the arc consistently gets bent downwards (see
Fig.\ref{fig:flarelab}).
This fact has been reported by Bellan et al. \cite{bib:bellan2001} as well.
In the simulations, we observe the same behavior in run 3 (constant
Alfv\'en velocity) and we explain it by the fact that the mass density
$\propto B^2$ assumes its largest values in the apex of the
plasma arc.
The reason for this is that the initial dipole field $\vec{B}_d$ is
weakest there which leads to a relatively strong pinching effect
and hence large values of $\vec{B}$.
The large mass density close to the apex causes
that part of the arc to lag behind during the arc expansion, leading to the
characteristic dip.

We repeated the simulation of case 3 with an increased value for the
viscosity ($\nu=0.1$ instead of the previous $\nu=0.05$) and achieved
an even better agreement with the experimental pictures
(Fig. \ref{fig:comp} a).  With this modification, the overall structure of
the arc appears smoother compared to the previous run
3.
The data of this high viscosity run has been used to produce a
perspective view of the arc structure, showing magnetic field lines and
electric current lines (Fig. \ref{fig:comp} b). In this view, the
overall structure of the arc as a helically distorted current channel
becomes much more evident than from the axes-parallel views and shows
an excellent agreement also with an image published in
\cite{bib:bellan2001}.

Taking these findings together, we conclude that the present
simulations represent a successful initial step to model the overall
dynamics of the experiment in the framework of ideal MHD.
Despite the fact that the present models contain a number of ad-hoc
assumptions (like a specific density model, the presence of viscosity
and a simple isothermal energy transport equation), the results are
remarkably encouraging.
For the future, we expect improved experimental diagnostics to give
important additional information that in turn might lead to a
more adequate theoretical and numerical description and hence to an
even better agreement between both approaches.

\subsection*{Acknowledgments}
Access to the JUMP multiprocessor computer at the FZ J\"ulich was made
available through project HBO18. Part of the computations were
performed on a Linux-Opteron cluster supported by HBFG-108-291.  This
work benefited from partial support through DFG-SO 380, DFG-GK 1051
and VH-VI-123.

\end{document}